\documentclass[twoside,a4paper,11pt]{sea10}
\usepackage{graphicx}
\usepackage{hyperref}
\usepackage{movie15}
\begin{document}
\pagenumbering{arabic}
\pagestyle{myheadings}
\thispagestyle{empty}
{\flushleft\includegraphics[width=\textwidth,bb=58 650 590 680]{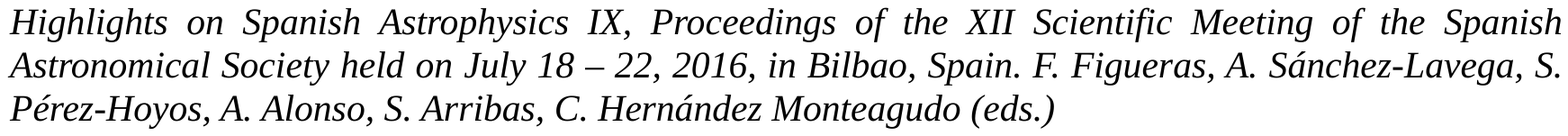}}
\vspace*{0.2cm}
\begin{flushleft}
{\bf {\LARGE 
%
GALACTIC CHEMICAL EVOLUTION
%
}\\
\vspace*{1cm}
%
M. Moll{\'a}$^{1}$,
O. Cavichia$^{2}$, 
R. da Costa$^{3}$,
B.K. Gibson$^{4}$,
and 
A.I. D{\'{\i}}az$^{5,6}$
%
}\\
\vspace*{0.5cm}
%
$^{1}$ Departamento de Investigaci\'{o}n B\'{a}sica, CIEMAT, Avda. Complutense 40, E-28040 Madrid. (Spain)\\ 
$^{2}$ Instituto de F\'{i}sica e Qu\'{i}mica, Universidade Federal de Itajub\'{a}, Av. BPS, 1303, 37500-903, Itajub\'{a}-MG, Brazil\\
$^{3}$ Instituto de Astronomia, Geofisica e Ci{\^e}ncias Atmosf{\' e}ricas, Universidade de S{\^a}o Paulo, 05508-900, S{\^a}o Paulo-SP, Brazil\\
$^{4}$ E.A Milne Centre for Astrophysics, University of Hull, HU6~7RX, 
United Kingdom\\
$^{5}$ Universidad Aut\'{o}noma de Madrid, 28049, Madrid, Spain \\
$^{6}$ Astro-UAM, Unidad Asociada CSIC, Universidad Aut{\'o}noma de Madrid,28049, Madrid, Spain
%
\end{flushleft}
%
\markboth{
 Galactic Chemical Evolution
}{ 
%
Moll{\'a} et al.
%
}
\thispagestyle{empty}
\vspace*{0.4cm}
\begin{minipage}[l]{0.09\textwidth}
\ 
\end{minipage}
\begin{minipage}[r]{0.9\textwidth}
\vspace{1cm}
\section*{Abstract}{\small
We analyze the evolution of oxygen abundance radial gradients resulting 
from our chemical evolution models calculated with different 
prescriptions for the star formation rate (SFR) and for the gas infall 
rate, in order to assess their respective roles in shaping gradients. We 
also compare with cosmological simulations and confront all with recent 
observational datasets, in particular with abundances inferred from 
planetary nebulae.  We demonstrate the critical importance in isolating 
the specific radial range over which a gradient is measured, in order 
for their temporal evolution to be useful indicators of disk growth with 
redshift.
%
\normalsize}
\end{minipage}
\section{Introduction \label{intro}}
Chemical elements appear in the Universe as a consequence of three 
production scenarios: 1) Big Bang Nucleosynthesis; 2) Fragmentation 
processes (Cosmic Rays); and 3) Stellar Nucleosynthesis.  H disappears 
at the same time that metals (elements heavier than He) increase their 
abundances. Elements are created, eventually ejected and diluted by the 
interstellar medium, and ultimately incorporated into successive 
generations of stars. If the process of star formation occurs rapidly, 
then the increase of the elemental abundances is also rapid; if it is 
slow, the abundances also increase slowly. In this way, abundance 
patterns give clues regarding star formation: when, how, and with which 
rate, stars form. Galactic Chemical Evolution (GCE) models try to 
explain how and when the elements appear based upon some hypothesis 
\cite{clay87,tin78}.  In the process of understanding the creation and 
evolution of the elements, we might compare model predictions with data 
and deduce the evolutionary histories of galaxies.

One of the best known features of spiral galaxies is the existence of a 
radial gradient of abundances in their disks \cite{hw99}. This gradient 
was first observed in the Milky Way Galaxy (MWG) and then in other 
external galaxies \cite{sha83,mccall85,zar94}, and it is now well 
characterized in our local Universe \cite{sanchez14}. The radial 
gradient was interpreted as due to differences in the star formation 
rate or the gas infall rate along the disk, although other reasons are 
also possible --see \cite{gotz92} for a detailed review--. In order to 
understand the role of the involved processes, numerical chemical 
evolution models were early developed \cite{lf83,lf85,dt84,td85,mf89,fer92,fer94}. 
Most of these models were able to reproduce the present state of our 
Galaxy, however not all of them predict the same evolution with time. 
For example, models from \cite{dt84,mol90} predict an initially flat 
radial distribution, which then steepens with time. Conversely, models 
by \cite{fer92,fer94} possess a steep initial radial abundance gradient 
which flattens with time. The first scenario is the consequence of the 
existence of the disk before starting the evolution (i.e., initial 
conditions), combined with an infall of primordial gas which dilutes the 
metal enrichment of the disk, preferentially in the outer regions. In 
the second scenario, however, the infall of gas forms out the disk and, 
furthermore, it contributes to the metal budget because the infalling 
gas is not primordial: there is also star formation in the halo before 
the gas falls onto the disk. Therefore, it seems that the radial 
gradient of abundances, observed in spirals is dependent upon the 
scenario of formation and evolution of the disk --see \cite{tosi96} and 
references therein. In fact, as \cite{kop94} explained, the radial 
gradient may be only modified by inflows or outflows of gas, and, 
besides the possible variations of the initial mass function (IMF) or 
stellar yields along the galactocentric radius, the only way to change a 
radial gradient of abundances is to have a star formation rate (SFR) or 
an infall rate changing with galactocentric radius. Most of models able 
to reproduce the present state of the solar neighborhood and the 
Galactic disk as a whole follow one of two trends: a) the gradient is 
steep, and then it flattens with time 
\cite{fer94,pran95,mol97,por00,hou00}, or, b) it is flat initially and 
it then steepens with time \cite{td85,mol90,chia97}. Besides this 
question of the scenario for each chemical evolution model, radial flows 
of gas or stellar migration may also modify, flattening, the radial 
distributions of abundances for stars of different ages --see 
\cite{kub15} for a review about this subject.

From an observational perspective, the question of appropriate datasets 
against which to compare has been debated vigorously. In \cite{mol97} we 
compared the chemical evolution results with two types of observational 
data: planetary nebulae (PNe) of different masses (that is, different 
ages) to oxygen abundance radial distributions at different times, and 
open and globular clusters, to estimate the global metallicity in 
objects of different ages. The problem with PNe is that to estimate 
their galactocentric distances and their stellar masses (and, therefore, 
their ages) is a process with large uncertainties. The open cluster age 
determination, in turn, is not an easy task, too, although the error 
bars are smaller in this case. However the box enclosing the data 
corresponding to these objects basically falls in the same region of the 
young stars in a plot of [Fe/H] {\sl vs} radius. In the end, the most 
accurate date are those of globular clusters, and based mainly on their 
metallicities, we reached the conclusion that our models with a 
flattening of the radial gradient of abundances were good enough to 
reproduce the observational data. This situation has changed, however, 
in the light of the new observational results of stellar metallicities 
and of the PNe determinations of age and distance, as we will show 
later.

\begin{figure}[h]
\center
\includegraphics[scale=0.6]{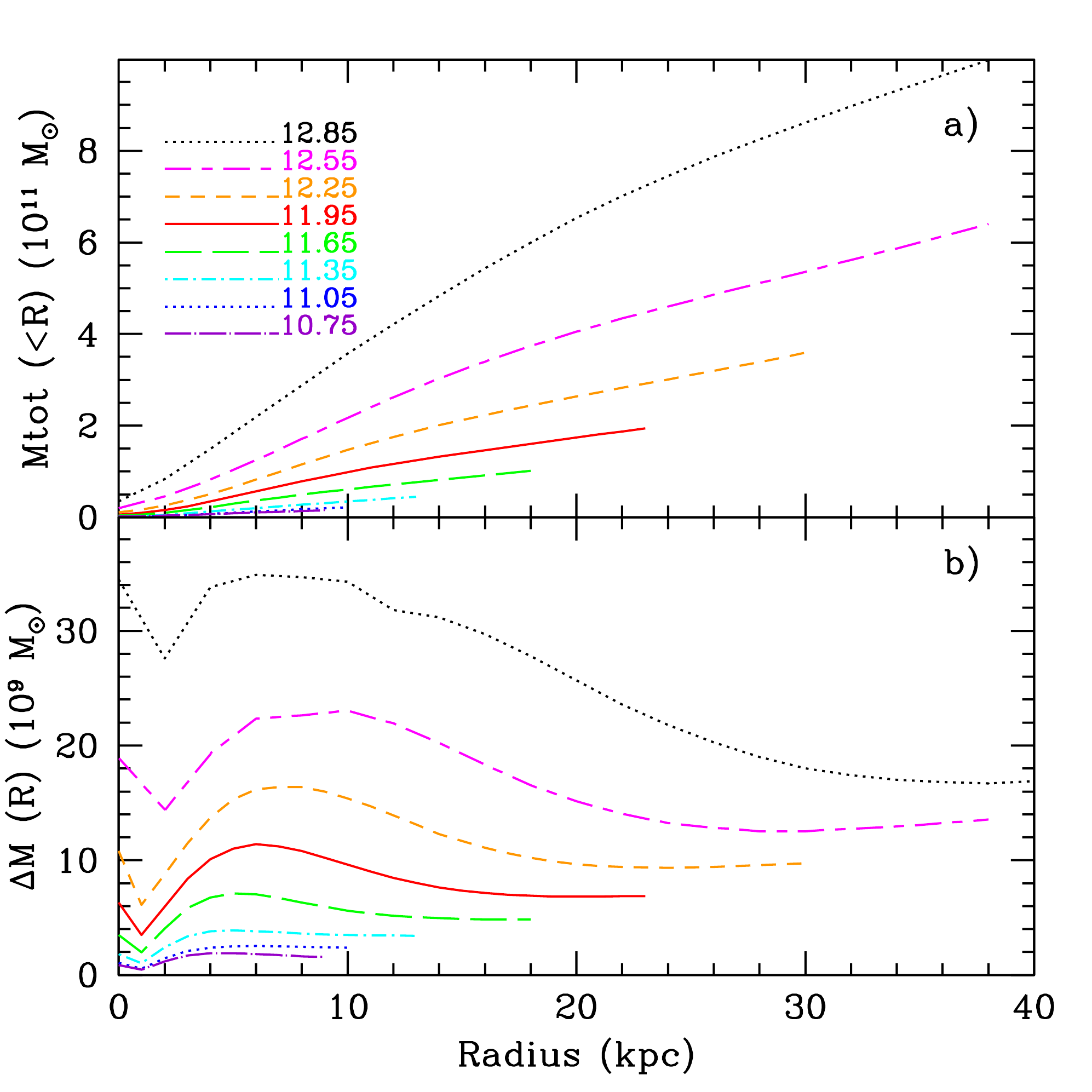} 
\caption{\label{dis} Radial distributions of: a) the total mass; and b) 
the mass enclosed in each radial region. Different lines correspond to 
different virial mass values, as labelled. Taken from \cite{mol16}.
}
\end{figure}
We, therefore, think that it is time to revise this question of the 
evolution of the radial gradient of abundances. To do so, we have 
generated a new suite of chemical evolution models for which we have 
revised all the necessary inputs for our code.  Here, we will analyze 
the results corresponding to a Milky Way Galaxy (MWG) like model.

\section{Chemical Evolution Models}
In a chemical evolution model, a scenario is assumed in which there is a 
given mass of gas in a certain geometric region. This mass is converted 
to stars by following an assumed star formation law. A mass ejection 
rate appears as a consequence of the death of stars. Often, some 
hypothesis concerning gas infall and outflows are included. The ejected 
mass depends, therefore, on the remnant of each stellar mass, on the 
mean-lifetimes of stars and on the IMF employed. This suite of models, 
based on \cite{fer92,fer94}, are an update of those from \cite{mol05}, 
hereinafter MD05. We start with a mass in a proto-halo which is 
calculated from \cite{sal07}.

We have computed 16 radial distributions within a range of virial masses 
$[5 \times10^{10}-10^{13}]M_{\odot}$, which implies maximum rotation 
velocities in the range $[42-320]$ km\,s$^{-1}$ and leads to disks with 
total masses in the range $[1.25\times10^{8}-5.3\times 10^{11}]M_{\odot}$. We 
show in Fig.~\ref{dis} the radial distributions we have obtained for 
various virial masses, as labelled.

Shaping the predicted radial abundance patterns are also the: 1) stellar 
yields (with the corresponding mean-lifetimes of stars) and the IMF; 2) 
infall rate of gas over the disk, and 3) star formation law, as outlined 
next.

\subsection{Stellar Yields} 
\begin{figure}[t]
\center
\includegraphics[scale=0.6,angle=0]{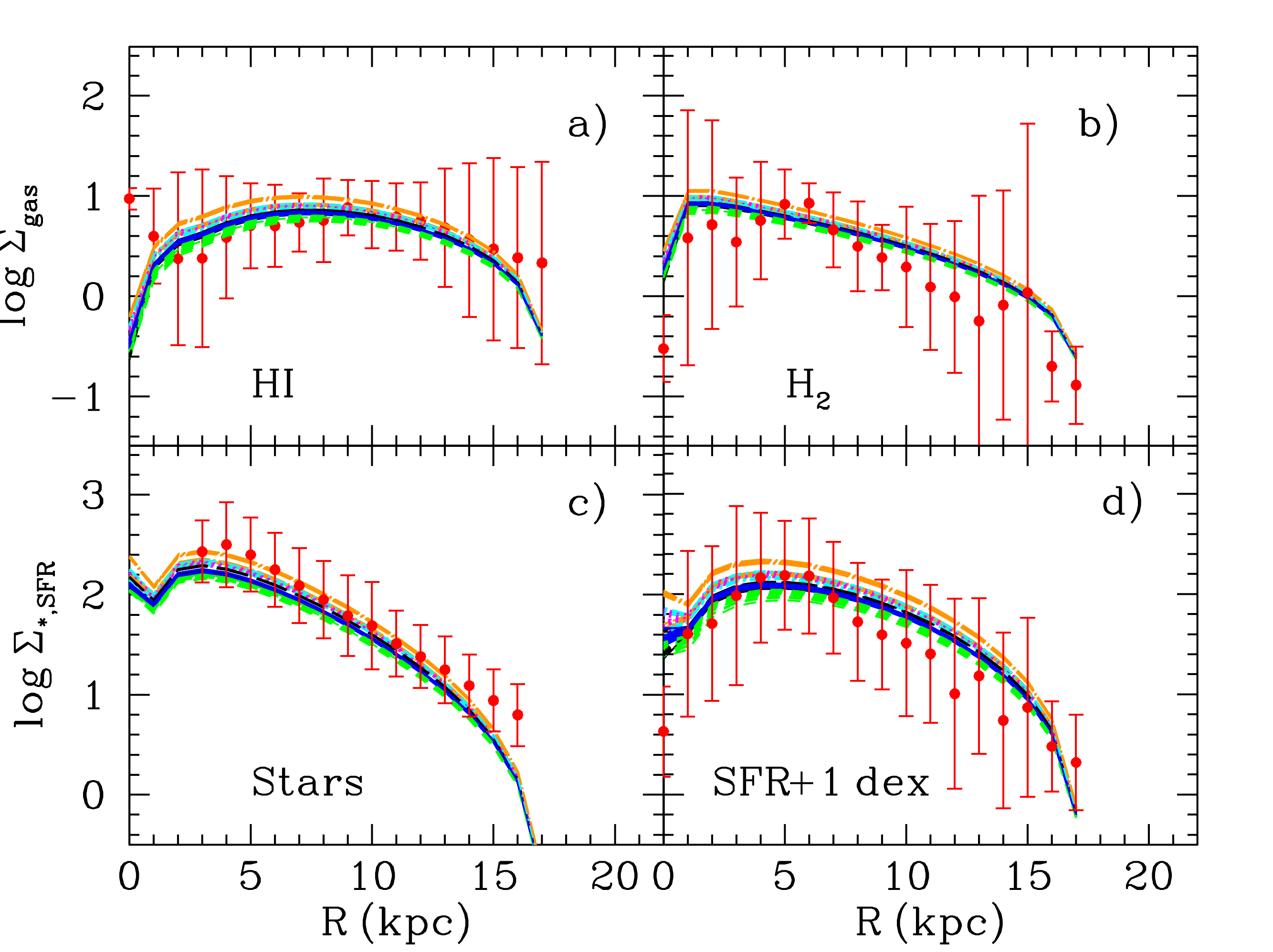} 
\caption{\label{profiles} Radial distributions of surface density of: 
a) diffuse gas, HI; b) molecular gas, H$_2$; c) stars and d) SFR. 
Different lines correspond to different IMF $+$ stellar yields 
combinations. Dots with error bars are the data for MWG. Taken from 
\cite{mol15}.
}
\end{figure}
We have divided, as usual, the stellar masses into two ranges 
corresponding to low- and intermediate-mass stars ($m < 8 M_{\odot}$), 
and massive stars ($m > 8 M_{\odot}$). The first eject mainly He$^{4}$, 
C$^{12,13}$ and $N^{14,15}$; a small amount of O can be generated in 
some yield prescriptions, as well as various s-process isotopes.  
Massive stars, in turn, produce C, O, and all the so-called 
$\alpha$-elements, up to Fe.  The literature of stellar yield generation 
is a rich one, with various works differing from one another due to the 
intrinsically different input physics to the underlying stellar models.

\begin{figure}[t]
\center
\includegraphics[scale=0.35]{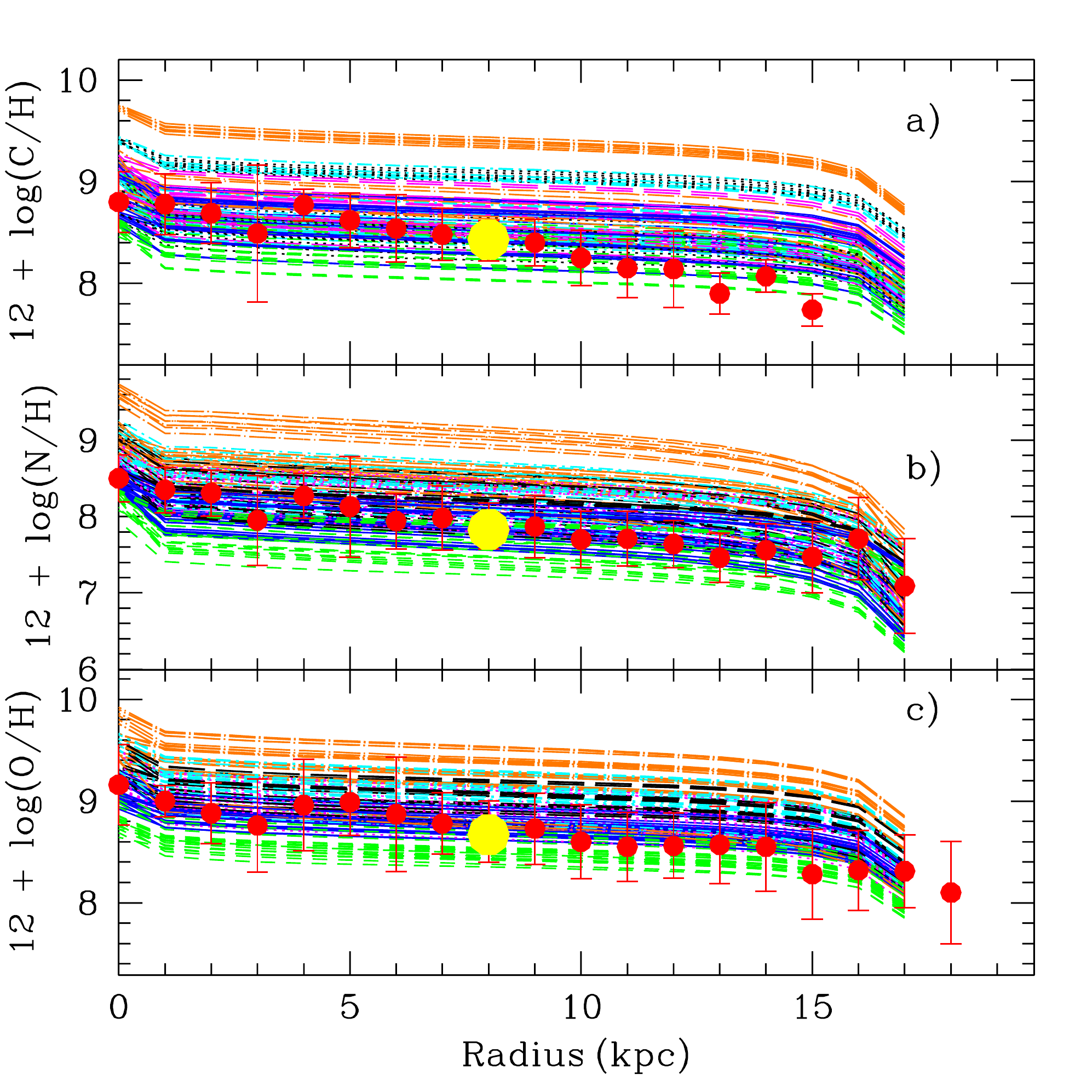} 
\includegraphics[scale=0.35]{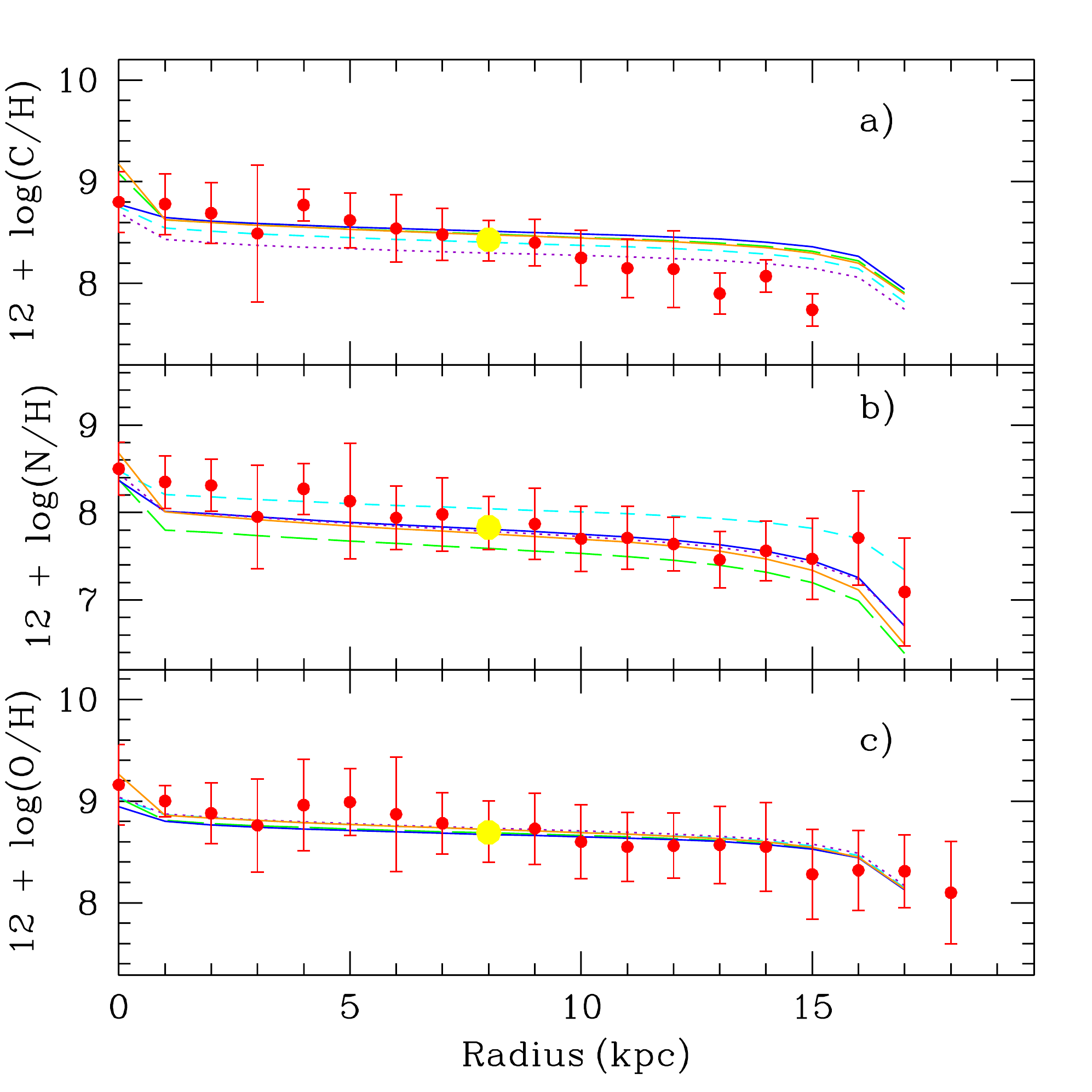} 
\caption{\label{abun} Radial distributions of C, N and O elemental 
abundances for: a) models with 144 IMF $+$ stellar yield combinations; 
b) the models selected as the best. Different lines correspond to 
different IMF $+$ stellar yields combinations. Dots with error bars are 
the data for MWG as taken from the compilation in \cite{mol15}.  The 
yellow large dot in each panel is the solar abundance for the 
corresponding elements. Taken from \cite{mol15}.
}
\end{figure}

Furthermore, it is necessary to deconvolve these stellar yields with the 
IMF. Again, there are in the literature a certain number of different 
expressions to define this IMF. Therefore, we have addressed this 
question in a work described in \cite{mol15}. There we have computed a 
chemical evolution model for MWG by using 6 different stellar yields for 
massive stars, 4 different yields sets for low and intermediate mass 
stars, and 6 different IMFs. Analyzing these 144 permutations and 
comparing their results with the observational data for our Galaxy disk, 
we determined which combinations are most valid in reproducing the data. 
Following these results, the different IMF $+$ stellar yields 
combinations produce basically the same radial distributions for gas 
(diffuse and molecular), and stellar $+$ star formation surface 
densities, as shown in Fig.~\ref{profiles}. The corresponding radial 
distributions for the elemental abundances of C, N and O, as show in 
Fig.\ref{abun}, left panel, are very different in their absolute values, 
with some far from the observational data, while others lie closer to 
the data. They do, however, show a similar slope for these radial 
distributions, suggesting that the selection of one or other combination 
is not particularly useful in modifying the radial abundance \it 
gradient\rm. In any case, there only 8 combinations IMF $+$ stellar 
yields able to reproduce the MWG data with a high probability ($P> 
97$\%). We use from here the yield combination given by 
\cite{gav05,gav06} and \cite{lim03,chi04}, with the IMF of \cite{kro02}, 
which provides a good match to the data, as we show in Fig.\ref{abun}, 
right panel. See more details in \cite{mol15}.

\subsection{Infall rate}
In our scenario, the gas initially in the proto-halo falls to the 
equatorial plane where the disk forms. In our old MD05 models, we 
assumed that this infall rate depends on the total mass of each galaxy, 
$M$, through a collapse time-scale $\tau$, defined by the expression 
$\tau \propto M^{-1/2}$, from \cite{gal94}. Moreover, within the disk, 
and following the same idea as other authors, we had assumed that $\tau$ 
has a radial dependence, being shorter in the inner regions of the disk 
and longer in the outer ones. In fact, we used an exponential function 
to compute this time scale, at variance with other models where a linear 
relationship with radius was used.  Now, we profit from the new 
knowledge about the relationship between the halo mass and the 
corresponding mass of the associated disks to estimate which must be the 
infall rate from the halo to obtain the correct disk. By using the 
equation from \cite{shan06}:
\begin{equation} 
M_{\textrm{\scriptsize star}}\,[{\rm M_{\odot}}]=2.3\,\times 10^{10} \frac{\left(M_{\textrm{\scriptsize vir}}/3.10^{11}\,
{\rm \mbox{M}_{\odot}}\right)^{3.1}}{1+\left(M_{\textrm{\scriptsize vir}}/3.10^{11}\,{\rm \mbox{M}_{\odot}}\right)^{2.2}},
\label{shankar06}
\end{equation}
\noindent
it is possible to determine the appropriate final disk mass. 
From this expression and using the rotation curves and equations from 
\cite{sal07}, we obtain the radial distributions of mass in the 
proto-halo and in the disk in each geometrical region at a given 
galactocentric distance; this is defined by a cylinder, as given in 
Fig.\ref{dis}. From these masses we obtain the collapse time in each 
radius $R$ as:
\begin{equation}
\tau(R)=-\frac{13.2}{\ln{\left(1-\frac{\Delta M_{\textrm{\scriptsize D}}(R)}{\Delta M_{\textrm{\scriptsize tot}}(R)}\right)}}\,[\mbox{Gyr}]
\end{equation}

The collapse time scale coming from this expression for a MWG-like model 
is smoother than the old exponential function used in MD05. The 
resulting infall rates produced by this collapse timescale are thus 
different than before: they show a smooth evolution for disks and 
stronger for bulges. The infall rates for different mass disks show 
variations only in the absolute values, with a very similar behavior for 
all galactic masses. The infall rate is very low in the outer regions of 
disk, and maintains that with redshift.  This implies that
the SFR is also low for all times. A sharp break in the disk is associated
with the dramatic decline in star formation at a given radius.

When we compare these results with the ones coming from cosmological 
simulations we see that our older models possessed stronger evolution 
with time, appearing more similar to simulations associated with 
spheroids/early-types; our newer models are more akin to simulations of 
late-type disks of the correct size and mass, as shown in Figs. 3 and 6 
of \cite{mol16}.

\subsection{Star formation law}
The last of our ingredients refers to the SFR law. In the halo, we 
assume a Schmidt law to form stars with an index $n=1$. However, in the 
disc we assume that the SFR takes place in two steps: first, molecular 
clouds form from the diffuse gas. Then stars form from cloud-cloud 
collision or by the interaction of the molecular clouds with massive 
stars. The first process in the halo depends on the volume with an 
efficiency taken as constant for all haloes. Conversely, the second 
process is a local one and also assumed to possess a constant 
efficiency. In our older MD05 models, we had assumed that the other two 
processes were dependent on the volume of the disk and on an efficiency 
considered to be a free parameter. Both of these latter efficiencies, 
$\epsilon_{c}$ and $\epsilon_{s}$, for the cloud and star formation, 
respectively, were varied simultaneously through a number $T$ which 
defined the set ($\epsilon_{c}$, $\epsilon_{s}$). In this new grid of 
models we have included new prescriptions to form molecular gas from 
diffuse gas, thus eliminating the efficiency $\epsilon_{c}$ as a free 
parameter. Finally only the efficiency $\epsilon_{s}$ remains as a free 
parameter.

\begin{figure}[h]
\center
\includegraphics[scale=0.35]{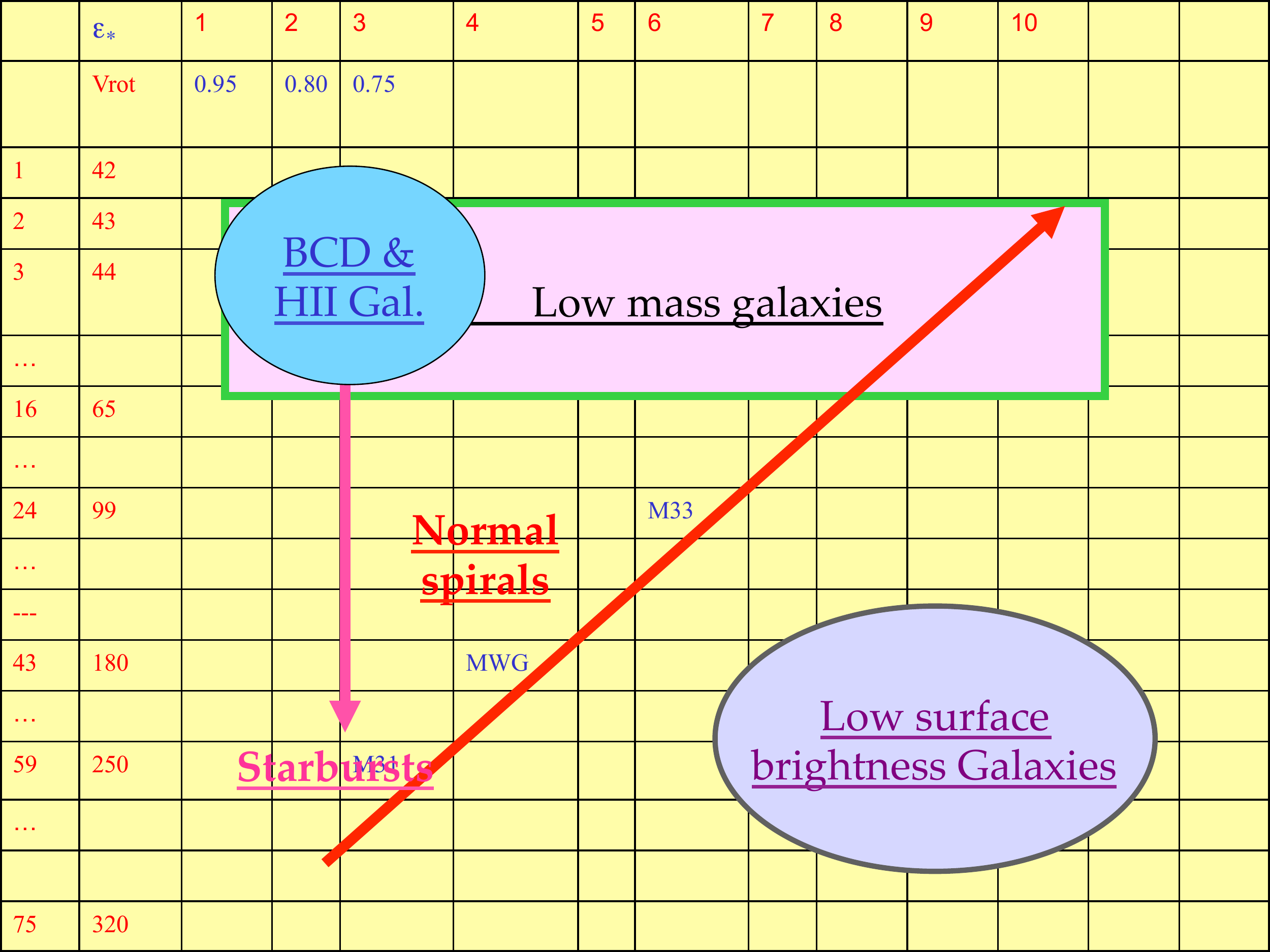} 
\caption{\label{scheme} Scheme of our bi-parametric suite of models.
}
\end{figure}
With these inputs we have finally a bi-parametric suite of models, as 
shown in Fig.\ref{scheme}, defined by the virial mass (which implies the 
size of the disk and the collapse timescale) and the efficiency to form 
stars $\epsilon_{s}$.  These models are valid for low mass galaxies as 
well as massive ones. The starburst or H{\sc ii} galaxies are included 
within the low mass and high efficiencies models, while low surface 
brightness galaxies will be among the more massive ones with low 
efficiencies. The MWG model  corresponds to $\log{M_{\textrm{vir}}}=11.95$ and an 
efficiency $\epsilon_{s}(T=4)$.

\section{Results}
\begin{figure}
\center
\includegraphics[scale=0.28,angle=0]{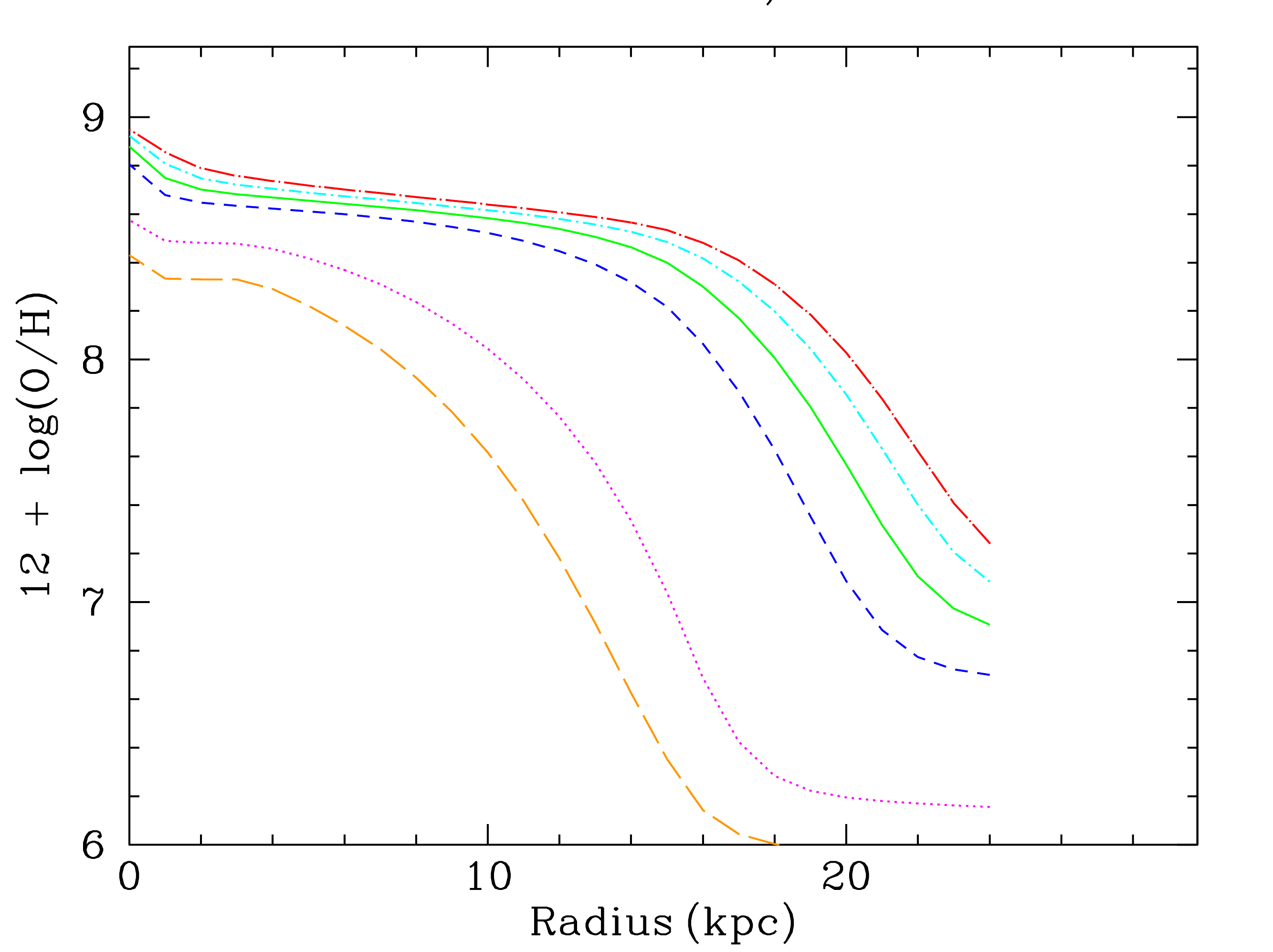} 
\includegraphics[scale=0.28,angle=0]{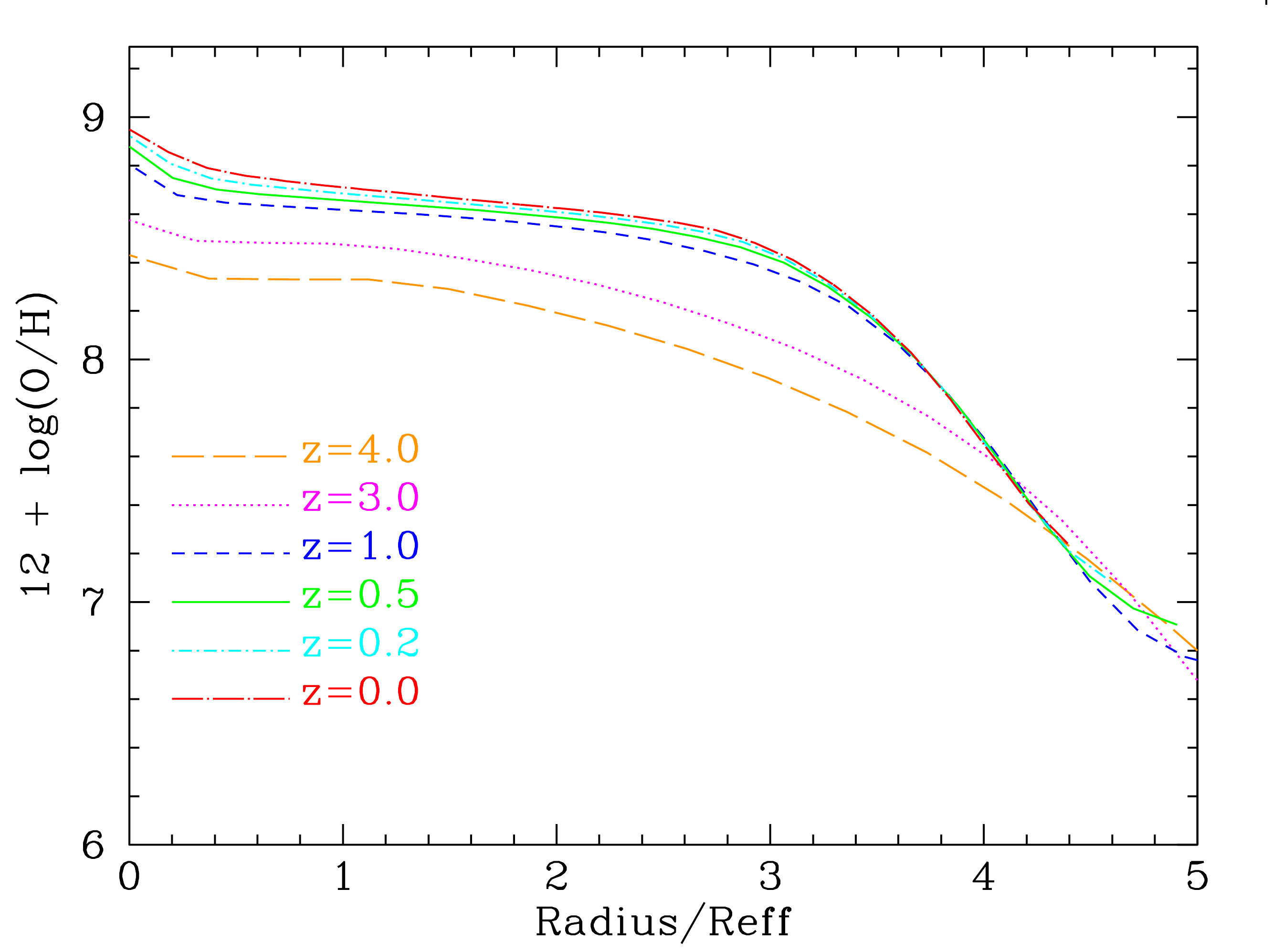} 
\caption{\label{oh}Radial distribution for the oxygen abundance, 
12+log(O/H), for 5 different redshifts: z=0, 0.5, 1, 2, 3 and 4 with 
red, cyan, green, blue, magenta and orange lines, respectively, as a 
function of: left) $R$ in $kpc$, and right) as a function of the 
normalized radius $R/R_{\textrm{eff}}$.
}
\end{figure}

One of the most important results derived from this work for our MWG 
model is the time evolution of the radial abundance gradient, in 
particular, for this work, oxygen. In Fig.\ref{oh}, left panel, we 
represent the oxygen radial distribution, as $12+\log(O/H)$ {\sl vs} $R$ 
for 5 different values of redshifts (see caption). The first thing we 
note is that it is impossible to distinguish the results among $z=0$ and 
$z<2$ when the radial range is restricted to $R\sim 14$\,kpc, that is 
around the optical radius or slightly larger. Only for higher redshifts 
is it possible to see a difference.

This implies that a clear separation in data coming from H{\sc ii} 
regions and PNe of 5-8\,Gyr old will not be apparent, such as 
\cite{mag16} find. The second fact we see here is that the radial 
gradient is the same for all redshift, when we measure it within the 
optical radius (which, obviously, decreases with increasing redshift). 
This way, for $z=4$ the radial gradient is the same as for $z=0$ when 
measured out only to $R\sim 5$\,kpc instead of (say) out to 15 kpc as in 
the case for redshift $z$=0. This is quite reasonable since beyond the 
optical radius there are few stars present to ionize the gas, and 
therefore deriving the gradient from the outer disk is not relevant for 
comparisons with empirical data.  This fact is more clearly shown in the 
right panel of the same Fig.\ref{oh}, where $12+\log(O/H)$ is represented 
as a function of the normalized radius $R/R_{\textrm{eff}}$.

\begin{figure}
\center
\includegraphics[scale=0.50,angle=0]{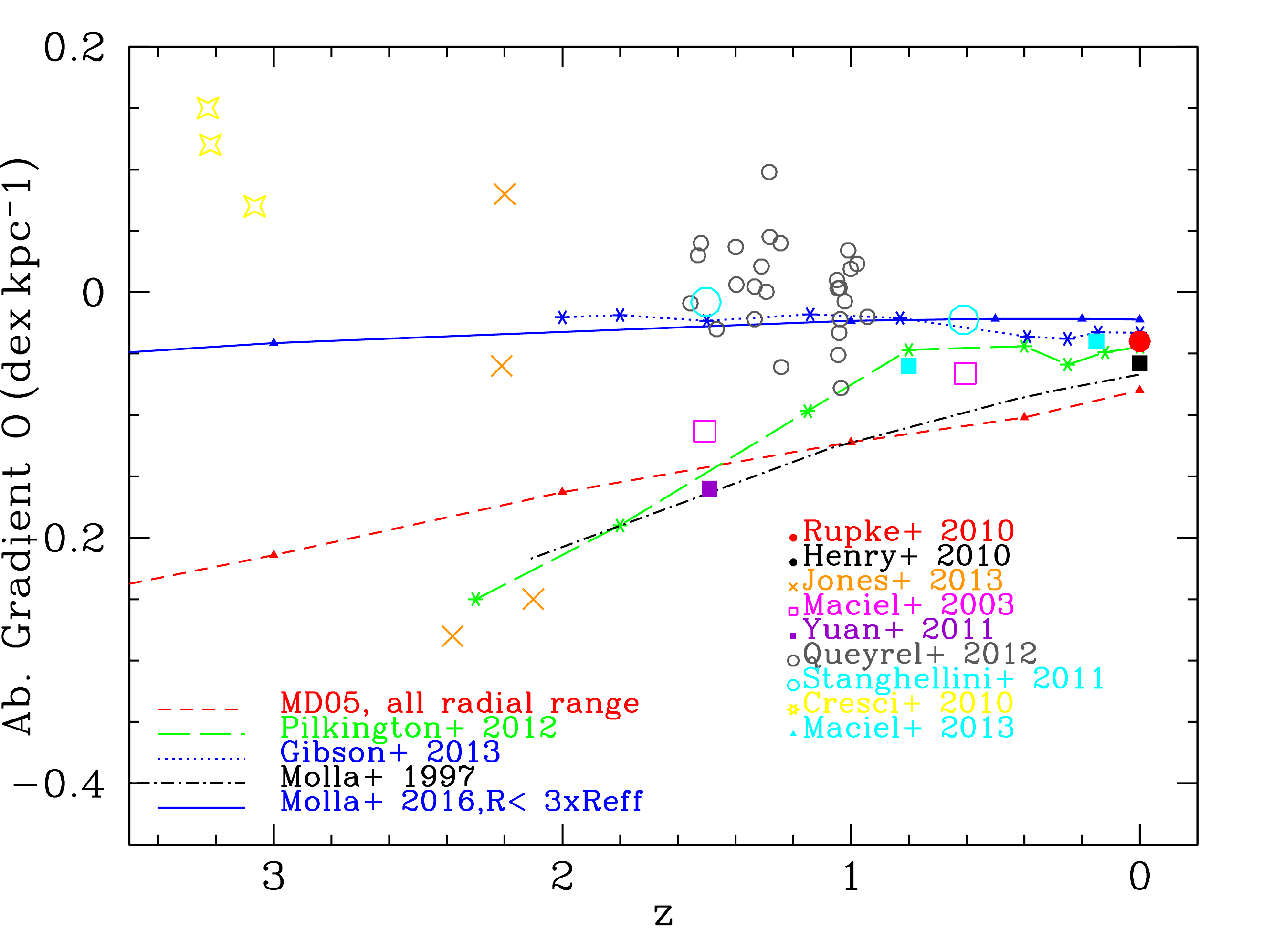} 
\caption{\label{grad}Radial gradient for the oxygen abundance, as 
$\textrm{dex}\,\textrm{kpc}^{-1}$ as a function of redshift, compared with our old model
predictions, with other data (see  text), as labelled, and with cosmological simulations results from 
\cite{pil12,gib13}.}
\end{figure}

We have computed the temporal evolution of the radial oxygen gradient 
for the MWG model.  To do so, we have fit a least-squares line to our 
results, restricting the fit to radii $R< 3R_{\textrm{eff}}$. The resulting 
evolution is shown by the blue line in Fig.\ref{grad}. In that figure we 
have also plotted some observational data from MWG 
\cite{mac03,rup10,hen10,stan10,mac13,gen14,xiang15,cun16,anders16}, and 
also from the other spiral galaxies \cite{yuan11,quey12, jon13,mag16}. 
We see that our new results for a MWG-like model are in better agreement 
with the most recent data, particular data derived from PNe.  The predictions from
cosmological simulations from \cite{pil12} and \cite{gib13} are also plotted, showing
that our new model also agrees with the last one in which the feedback of the
star formation treatment is included. Much work 
remains to be done to explain the diversity of gradients beyond that of 
the MWG though; this is left to a more detailed future study.

\section{Conclusions}
\begin{enumerate}
\item A grid of chemical evolution models with 16 different total dynamical masses in the range 10$^{10}$ to 10$^{13}$ M$_{\odot}$ is calculated.
\item 10 values of efficiencies $\epsilon_{s}$ to form stars from molecular clouds  are used ($0 < \epsilon_{s} < 1$).
\item The best combination with an IMF from \cite{kro02}$+$ stellar yields from \cite{gav05,gav06}  $+$ \cite{lim03,chi04}, is used.
\item The stellar yields $+$ IMF do not change the radial distributions of disks.
\item Using \cite{shan06} prescriptions for $M_{\textrm{halo}}/M_{\textrm{disk}}$, we obtain the necessary infall rates to reproduce the radial profiles of  galaxy disks
\item The HI-to-H$_2$ formation prescriptions from Ascas{\'\i}bar \& Romero (in prep.) seem the best ones to reproduce the MWG disk.
\item The slope of the Oxygen abundance radial gradient measured for $R<3\,R_{\textrm{eff}}$ has a value $-$0.06 dex\,kpc$^{-1}$, or -0.12 dex\,$R_{\textrm{eff}}^{-1}$, when measured versus a normalized radius, in excellent agreement with the universal radial gradient obtained for CALIFA galaxies \cite{sanchez14}.
\item The same slope is obtained for all efficiencies and all galaxy masses.
\item This slope does not change much with redshift when the infall rate is as smooth as we have obtained recently, compared with old models with stronger evolution.
\end{enumerate}
%
%
\small  
%
\section*{Acknowledgments}   
%
This work has been supported by DGICYT grant AYA2013-47742-C4-4-P.  This work has been supported financially by grant
2012/22236-3 from the S\~{a}o Paulo Research Foundation (FAPESP). This work has made use of the computing facilities of the Laboratory of
Astroinformatics (IAG/USP, NAT/Unicsul), whose purchase was made possible by the Brazilian agency FAPESP (grant 2009/54006-4) and the
INCT-A. MM thanks the kind hospitality and wonderful welcome of the Jeremiah Horrocks Institute at the University of Central Lancashire, the E.A. Milne Centre for Astrophysics at the University of Hull, and the Instituto de Astronomia, Geof\'{\i}sica e Ci\^{e}ncias Atmosf\'{e}ricas in S\~{a}o Paulo (Brazil), where this work was partially done. 
%

%

\begin{thebibliography}{}
\small
%

\bibitem{anders16} Anders, F., Chiappini, C., Minchev, I., et al.\ 2016,  A\&A, submitted (arXiv:1608.04951) 

\bibitem{asp09}{Asplund, M., Grevesse, N., Sauval, A.~J., \& Scott, P.\ 2009, ARAA, 47, 481}

\bibitem{br92}{Belley, J., \& Roy, J.-R.\ 1992, ApJS, 78, 61} 

\bibitem{bli06}{Blitz, L., \& Rosolowsky, E.\ 2006, ARAA, 650, 933} 

\bibitem{bp00}{Boissier, S., \& Prantzos, N.\ 2000, \textit{MNRAS}, 312, 398} 

\bibitem{bp99}{Boissier, S., \& Prantzos, N.\ 1999, A\&A, 307, 857}

\bibitem{cav14}{Cavichia, O., Moll{\'a}, M., Costa, R.~D.~D., \& Maciel, W.~J.\ 2014, \textit{MNRAS}, 437, 3688} 

\bibitem{chia97}{Chiappini, C., Matteucci, F., \& Gratton, R.\ 1997, ApJ, 477, 765} 

\bibitem{chi04}{Chieffi A., Limongi M., 2004, ApJ, 608, 405}

\bibitem{clay87}{Clayton, D.~D.\ 1987, ApJ, 315, 451}

\bibitem{cresci10}{Cresci, G., Mannucci, F., Maiolino, R., et al.\ 2010, Nature, 467, 811 }

\bibitem{cun16}{Cunha, K., Frinchaboy, P.~M., Souto, D., et al.\ 2016, Astronomische Nachrichten, 337, 922} 

\bibitem{diaz89}{Diaz, A.~I.\ 1989, Evolutionary Phenomena in Galaxies, 377}
 
\bibitem{dt84}{Diaz A.~I., Tosi M., 1984, \textit{MNRAS}, 208, 365} 

\bibitem{fer92}{Ferrini F., Matteucci F., Pardi C., Penco U., 1992, ApJ, 387, 138}

\bibitem{fer94}{Ferrini F., Molla M., Pardi M.~C., Diaz A.~I., 1994, ApJ, 427, 745} 

\bibitem{gal94}{Gallagher J.~S., III, Hunter D.~A., Tutukov A.~V., 1984, ApJ, 284, 544}

\bibitem{gav05}{Gavil{\'a}n, M., Buell, J.~F., \& Moll{\'a}, M.\ 2005, A\&A, 432, 861} 

\bibitem{gav06}{Gavil{\'a}n M., Moll{\'a} M., Buell J.~F., 2006, A\&A, 450, 509} 

\bibitem{gib13}{Gibson, B.~K., Pilkington, K., Brook, C.~B., Stinson, G.~S., \& Bailin, J.\ 2013, A\&A, 554, A47} 

\bibitem{gen14}{ Genovali, K., Lemasle, B., Bono, G., et al.\ 2014, A\&A, 566, A37 }

\bibitem{gotz92}{Goetz M., Koeppen J., 1992, A\&A, 262, 455}

\bibitem{hw99}{Henry, R.~B.~C., \& Worthey, G.\ 1999, PASP, 111, 919} 

\bibitem{hen10}{Henry, R.~B.~C., Kwitter, K.~B., Jaskot, A.~E., et al.\ 2010, ApJ, 724, 748} 

\bibitem{hou00}{Hou, J.~L., Prantzos, N., \& Boissier, S.\ 2000, A\&A, 362, 921} 

\bibitem{iwa99}{Iwamoto, K., Brachwitz, F., Nomoto, K., et al.\ 1999, ApJS 125, 439} 

\bibitem{jon13}{Jones, T., Ellis, R.~S., Richard, J., \& Jullo, E.\ 2013, ApJ, 765, 48 }

\bibitem{kar10}{Karakas, A.~I.\ 2010, \textit{MNRAS}, 403, 1413} 

\bibitem{ken98}{Kennicutt, R.~C., Jr.\ 1998, ApJ, 498, 541} 

\bibitem{kop94}{Koeppen J., 1994, A\&A, 281, 26} 

\bibitem{kro02}{Kroupa P., 2002, Sci, 295, 82}
 
\bibitem{kub15}{Kubryk M., Prantzos N., Athanassoula E., 2015, A\&A, 580, A127 }

\bibitem{lf83}{Lacey, C.~G., \& Fall, S.~M.\ 1983, \textit{MNRAS}, 204, 791} 

\bibitem{lf85}{Lacey, C.~G., \& Fall, S.~M.\ 1985, ApJ, 290, 154 }

\bibitem{lim03}{Limongi M., Chieffi A., 2003, ApJ, 592, 404}

\bibitem{mac94}{Maciel W.~J., Koppen J., 1994, A\&A, 282, 436} 

\bibitem{mac03}Maciel, W.~J., Costa, R.~D.~D., \& Uchida, M.~M.~M.\ 2003, A\&A, 397, 667 

\bibitem{mac13}{Maciel, W.~J., \& Costa, R.~D.~D.\ 2013, RMxA\&Ap, 49, 333} 

\bibitem{mag09}{Magrini L., Sestito P., Randich S., \& Galli D.\ 2009, A\&A, 494, 95} 

\bibitem{mag16}{Magrini L., Coccato L., Stanghellini L., Casasola V., Galli D.\, 2016, A\&A, 588, A91 }

\bibitem{mf89}{Matteucci, F., \& Francois, P.\ 1989, \textit{MNRAS}, 239, 885} 
 
\bibitem{mccall85}{McCall, M.~L., Rybski, P.~M., \& Shields, G.~A.\ 1985, ApJS, 57, 1} 

\bibitem{mol90}{Moll{\'a} M., D{\'\i}az A.I., Tosi M., 1990,  in Chemical and dynamical evolution of Galaxies, eds. F. Ferrini, J. Franco, 
\& F. Matteucci, ETS Editrice (PISA), 577}

\bibitem{mol97}{Moll{\'a}, M., Ferrini, F., \& Diaz, A.~I.\ 1997, ApJ, 475, 519}

\bibitem{mol05}{Moll{\'a}, M. \& D{\'{\i}}az, A.I.\ 2005, \textit{MNRAS}, 358, 521}

\bibitem{mol15}{Moll{\'a}, M., Cavichia, O., Gavil{\'a}n, M., \& Gibson, B.~K.\ 2015, \textit{MNRAS}, 451, 3693}

\bibitem{mol16}{Moll{\'a}, M., D\'{\i}az A.~I., Gibson, B.~K.,  et al., 2016, \textit{MNRAS}, }

\bibitem{pil12}{Pilkington, K., Few, C.~G., Gibson, B.~K., et al.\ 2012, A\&A, 540, A56} 

\bibitem{por00}{Portinari, L., \& Chiosi, C.\ 2000, A\&A, 355, 929 }

\bibitem{pran95}{Prantzos, N., \& Aubert, O.\ 1995, A\&A, 302, 69 }

\bibitem{pra00}{Prantzos, N., \& Boissier, S.\ 2000, \textit{MNRAS}, 313, 338}

\bibitem{quey12}{Queyrel, J., Contini, T., Kissler-Patig, M., et al.\ 2012, A\&A, 539, A93} 

\bibitem{rlp00}{Ruiz-Lapuente P., Blinnikov S., Canal R.,  Mendez J., Sorokina E., Visco A., Walton N., 2000, MmSAI, 71, 435} 

\bibitem{rup10}{ Rupke, D.~S.~N., Kewley, L.~J., \& Chien, L.-H.\ 2010, ApJ, 723, 1255} 

\bibitem{sal07}{Salucci P., Lapi A., Tonini C., Gentile G., Yegorova I., Klein U., 2007, \textit{MNRAS}, 378, 41} 

\bibitem{san08}{Sancisi, R., Fraternali, F., Oosterloo, T., \& van der Hulst , T. A\&ARv, 15, 189}

\bibitem{sanchez14}{S{\'a}nchez, S.~F., Rosales-Ortega, F.~F., Iglesias-P{\'a}ramo, J., et al.\ 2014, A\&A 563, A49}

\bibitem{shan06}{Shankar F., Lapi A., Salucci P., De Zotti G., Danese L., 2006, ApJ, 643, 14}

\bibitem{sha83}{Shaver, P.~A., McGee, R.~X., Newton, L.~M., Danks, A.~C., \& Pottasch, S.~R.\ 1983, \textit{MNRAS}, 204, 53} 

\bibitem{stan10}{Stanghellini, L., \& Haywood, M.\ 2010, ApJ, 714, 1096} 


\bibitem{tin78}{Tinsley, B.~M., \& Larson, R.~B.\ 1978, ApJ, 221, 554 }

\bibitem{td85}{Tosi, M., \& Diaz, A.~I.\ 1985, \textit{MNRAS} 217, 571} 

\bibitem{tosi96}{Tosi M., 1996, in From Stars to Galaxies: The Impact of Stellar Physics on Galaxy Evolution, ASP Conference Series, eds. 
C. Leitherer, U. Fritze-von-Alvensleben, and J. Huchra, 98, 299}

\bibitem{xiang15}{Xiang, M.-S., Liu, X.-W., Yuan, H.-B., et al.\ 2015, Research in Astronomy and Astrophysics, 15, 1209 }

\bibitem{yuan11}{Yuan, T.-T., Kewley, L.~J., Swinbank, A.~M., Richard, J., \& Livermore, R.~C.\ 2011, ApJL, 732, L14}

\bibitem{zar94}{ Zaritsky, D., Kennicutt, R.~C., Jr., \& Huchra, J.~P.\ 1994, ApJ, 420, 87}


%
%
\end{thebibliography}
\end{document}